\documentclass[prb,twocolumn,amsmath,amssymb]{revtex4}
\usepackage{hyperref}
\usepackage{graphicx}
\begin{document}

\title{Barkhausen instabilities from labyrinthine magnetic domains}

\author{A. Benassi$^{1,2}$, S. Zapperi$^{2,3}$}
\affiliation{
$^1$  Centro S3, CNR Istituto Nanoscienze, Via Campi 213/A, 41125 Modena, Italy  \\
$^2$ CNR-IENI Istituto per l'energetica e le interfasi, Via Cozzi 53, 20125 Milano, Italy \\
$^3$ ISI Foundation, Viale S. Severo 65, 10133 Torino, Italy
}


\begin{abstract} 
Experimental investigations of the scaling behavior of Barkhausen avalanches in out-of-plane ferromagnetic films yield widely different results for the values of the critical exponents despite similar labyrinthine domain structures, suggesting that universality may not hold
for this class of materials. Analyzing a phase field model for magnetic reversal, we show that avalanche scaling is bounded by
characteristic lengthscales arising from the competition between dipolar forces and exchange interactions.
We compare our results with the experiments and find a good  qualitative and quantitative agreement, reconciling apparent contradictions. Finally, we make some prediction, amenable to experimental verification, on the dependence of the avalanches behavior from the film thickness and disorder.
\end{abstract}

\maketitle

\section{Introduction}
Understanding magnetic hysteresis is a fundamental theoretical problem with important 
practical implications for magnetic devices \cite{bertotti}.  Magnetic reversal is usually associated with
crackling noise ---the Barkhausen effect--- due to the interaction between magnetic domains
and structural disorder \cite{sethna01,durin06}. The statistical properties of the Barkhausen noise are characterized
by scaling laws, a signature of an undelying non-equilibrium critical behavior, suggesting that it should be possible  
to separate magnetic materials into well defined universality classes characterized by the
same values of the critical exponents and the same form of the scaling functions \cite{papanikolaou11}. This program
has been successfully carried over for bulk materials with relatively simple parallel domain structures
where it is possible to predict distinct universality classes depending on the sample microstructure (amorphous or polycrystalline)
\cite{durin00}. 

The situation is more complicated in thin films which can show complex domain
structures due to the interplay of magnetic anisotropies and dipolar interactions.
In films with in-plane anisotropy, such as the MnAs films studied in Ref. \onlinecite{ryu07}, the domain structure crosses over from zigzag to rough
as a function of the temperature and this change is reflected by a crossover in the universality class
of Barkhausen avalanche statistics \cite{ryu07}. The main effect of temperature in this material is to modify the
strength of dipolar interactions \cite{ryu07}, a relevant parameter for the critical behavior as confirmed
by numerical simulations \cite{mughal10}. The magnetic properties of films with out-of-plane anisotropy have been intensively investigated in recent years due to their potentially higher bit packing density \cite{andra}. 
These systems  typically display a labyrinthine domain structure as shown by 
high resolution magneto-optical techniques \cite{young} or magnetic force microscopy \cite{schwarz,liebmann}. 
These techniques have been used to follow magnetic reversal and measure Barkhausen avalanches,  
reporting apparently contradictory results for the critical exponent of the avalanche statistics.
Whether universality holds for these systems remains thus an intriguing open question.

Here, we use a phase field model describing magnetization dynamics in an out-of-plane film \cite{jagla1,jagla2} to show 
that the scaling behavior of the avalanche size distribution is severely limited by dipolar interactions that
set a characteristic lengthscale to the problem. The phase-field model overcomes the limitations of
the dipolar random-field Ising model \cite{magni00} which is plagued by lattice effects. At the same
time the model, owing to its scalar structure, is much easier to simulate than the full micromagnetic equations.
Our numerical results are in good agreement with experimental data and allow to clarify the origin of the observed discrepancies between
the exponents reported in the literature. 
This paper is organized as follows: in Sec II we illustrate the phase field model used for our calculations, in Sec. III we present the results of our calculation including hysteresis loops and avalanche statistics, in the last section we compare our results with available experimenatal data and we draw some conclusions. 
\section{Model}
\begin{figure*}
\centering
\includegraphics[width=16.0cm,angle=0]{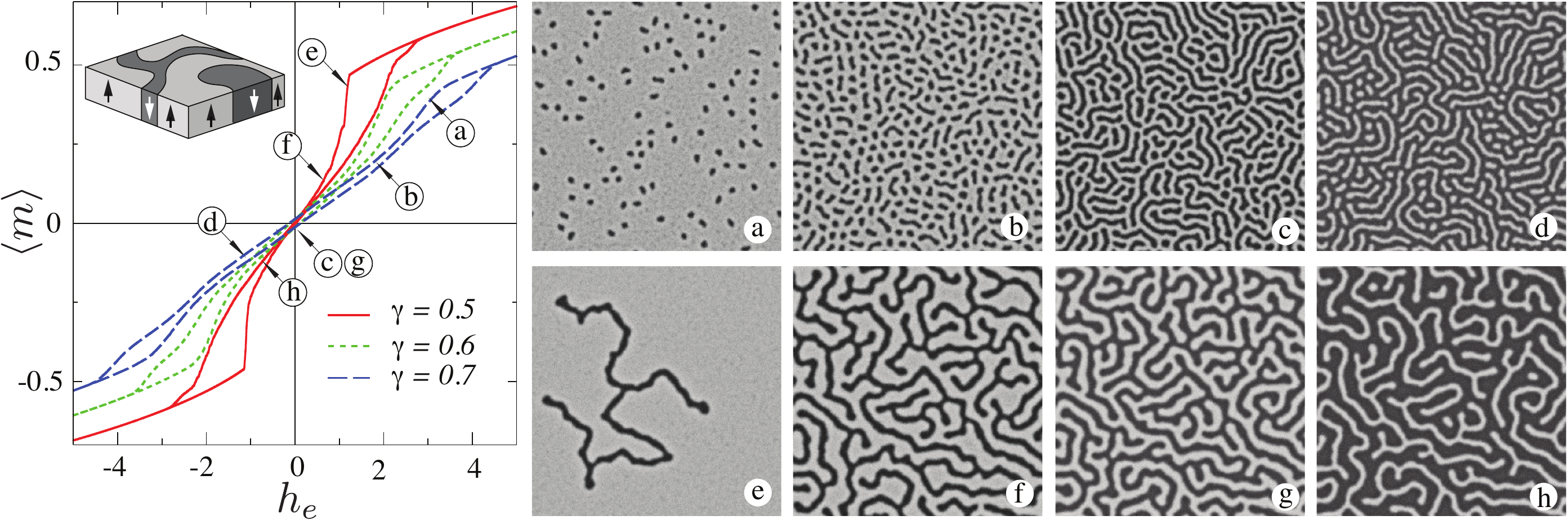}
\caption{(left) Hysteresis loops for different dipolar interaction strength $\gamma$. The inset sketches our model ferromagnetic thin film. Notice that the out-of-plane magnetization is constant along the film thickness. Panels (a-h) represent the film magnetization $m(x,y)$ for different external field values along the hysteresis loop for $\gamma=0.7$ (a-d) and $\gamma=0.5$ (e-h).}
\label{figura1}
\end{figure*}
In the model, the magnetization $\mathbf{M}(\mathbf{R})$ at position $\mathbf{R}$ is perpendicular to the film plane and constant along its thickness $d$, as illustrated in the inset of Fig.\ref{figura1}.
Therefore to describe the domains behavior, we adopt a two-dimensional phase field model where a dimensionless scalar order parameter $m({\mathbf R},t)=M_z({\mathbf R},t)/M_s$ represents the local magnetization value on the film plane, $M_s$ being the saturation magnetization \cite{jagla1}. The energy per unit thickness is given by:
\begin{widetext}
\begin{equation}
{\cal H}/d=\int d\mathbf{R} \bigg[V(m)+ \frac{A (\nabla m)^2}{2} -\frac{\mu_0 M_s^2 d}{4 \pi}\int d\mathbf{R}' \frac{m(\mathbf{R})m(\mathbf{R}')}{\vert \mathbf{R}-\mathbf{R}'\vert^3} -\mu_0 M_s m (H_{r}+H_{e})\bigg]
\label{hamilton}
\end{equation}
\end{widetext}
the first term mimics the anisotropy energy with a double well potential $V=-V_0(m^2/2-m^4/4)$ giving the order parameter two equivalent preferential orientations $\pm 1$, with the  constant $V_0$ determining the barrier height, in particular $V_0=K_u/4$ with $K_u$ uniaxial anisotropy constant of the material. The second term represents the exchange interaction, opposing any magnetization variation with strength $A$. The third term represents the long-range, 
non-local, dipolar interaction in the approximation of small film thickness, given a local value of $m$, it promotes the magnetization reversal in its surroundings ($\mu_0$ is the vacuum permeability). Pinning centers are modeled by a Gaussian uncorrelated random field $H_r$ with $\langle H_r({\mathbf R} )\rangle$=0 and $\langle H_r({\mathbf R} ) H_r({\mathbf R}' )\rangle=\Delta \delta({\mathbf R}- {\mathbf R}')$. Finally we 
consider a uniform external field $H_e$ that is slowly ramped up and down to produce an hysteresis loop.
We adopt a dimensionless scheme with unit length $\ell=\sqrt{A/V_0}$ and unit field $M_s$, defining ${\mathbf r}\equiv{\mathbf R}/\ell$, $h_e\equiv H_e/M_s$, $D \equiv \Delta/\ell^2M_s^2$ and $h_r \equiv H_r/M_s$. The order parameter equation of motion can be derived in the approximation of small time fluctuations, i.e. far from the critical point, through the hamiltonian functional derivative
$\partial M({\mathbf R},t)/\partial t=-\lambda \; \delta {\cal H}[M({\mathbf R},t)]/\delta M({\mathbf R},t)$ leading to:
\begin{equation}
\dot{m}=\alpha\bigg( \frac{d V}{d m}+\nabla^2 m \bigg) -\gamma \int d{\mathbf r}' \frac{m({\mathbf r}')}{\vert {\mathbf r} -{\mathbf r}' \vert^3}+h_{r}({\mathbf r})+h_{e}(t)
\label{one}
\end{equation}
$\alpha=V_0/\mu_0 M_s^2$ and $\gamma=d/\ell 4 \pi$ are dimensionless parameters and $\mu_0\lambda$ is the unit time.
For the numerical solution of eq. (\ref{one}) we used the finite-difference semi-implicit method of Ref. \onlinecite{jagla1}.
All the simulations have been performed keeping $\alpha=3$ and the disorder strength $D=2\cdot 10^{-3}$ except when another value is explicitly specified, the time step $\Delta t=0.5$ and the squared mesh step $\Delta x=\Delta y=0.4$.
\section{Results}
\subsection{Hysteresis loops and domain formation}
Typical hysteresis loops obtained by a slow ramp of the external field $h_e$ are shown in Fig.\ref{figura1} together with selected plots of the domain structures (see also supplemental movies 1 and 2). Depending on the strength of the dipolar term, tuned by the parameter $\gamma$, the magnetization dynamics can vary considerably.
For relatively large $\gamma$ (see Fig. \ref{figura1}(a-d)), we observe simultaneous nucleation of multiple droplet domains that grow as the external field is decreased below the saturation value. When the domain density is sufficiently high, domains 
merge giving rise to a labyrinthine structure. For smaller values of $\gamma$ (see Fig. \ref{figura1}(e-h))
a single nucleated droplet grows forming a labyrinth. This process occurs suddenly at a critical
value of the external field and corresponds to a jump in the hysteresis loop. Upon further decreases in the external field, domains expand by moving domain walls. It is remarkable that while the resulting labyrinthine domain structures are similar the dynamics that generates them is completely different. The two types of dynamics have been observed experimentally in irradiated and non-irradiated Ni films at low temperature \cite{pilet}.
\begin{figure}
\centering
\includegraphics[width=8.0cm,angle=0]{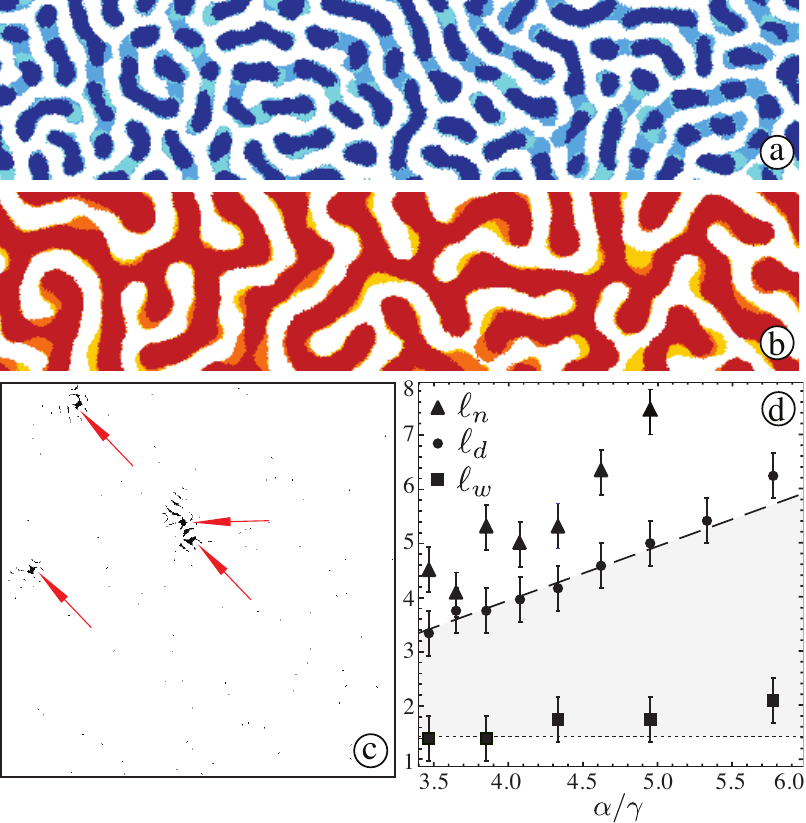}
\caption{Panels (a) and (b) show the evolution of the magnetization close to the center of the loop for $\gamma=0.7$ (a) and $\gamma=0.5$ (b), the color scale highlights the shape of the domains at three different subsequent times. Panel (c) describes the bridging phenomenon showing the  magnetization difference for two consecutive times. Four avalanches are highlighted having almost the same size, surrounded by smaller avalanches triggered by the bridging. Panel (d) shows the characteristic lengths in dimensionless units as a function of the ratio $\alpha/\gamma$. The dashed and dotted lines are the theoretical predictions: $\ell_d=\alpha/\gamma$ for the domain width  and $\ell_w=\sqrt{2}$ for the domain wall width.}
\label{figura2}
\end{figure}
Fig.\ref{figura2} (a) and (b) describe the two different domain dynamics with a color scale highlighting subsequent avalanches. Both pictures have been taken in the central part of the hysteresis loop at the same values of the applied field for $\gamma=0.7$ and $\gamma=0.5$ (see also supplemental movies 3 and 4). In Fig.\ref{figura2} (a), avalanches occur only close to the tails of the domain, leaving their width unchanged.  When two neighboring domains are close to each other, a single avalanche bridges them together, yielding
a characteristic avalanche lengthscale proportional to the domain width.

In Fig.\ref{figura2}(b) avalanches of a wide variety of sizes occur mostly by displacing domain walls. Domain bridging can be visualized by
 subtracting two subsequent images as shown in Fig.\ref{figura2}(c). There we see four bridging events having roughly the same area, together with
a swarm of smaller avalanches produced by the rearrangement of the surrounding domains. This avalanche triggering phenomenon is always present when the domain density is large.  
\subsection{Characteristic lengthscales}
Our simulations suggest the existence of an upper bound for avalanche sizes which appear to be limited by the width $\ell_d$ of the labyrinthine 
domains. We estimate the domain width by minimizing Eq.(\ref{hamilton}) with respect to $\ell_d$ assuming a simple form for
the magnetization field. In particular, we consider stripe domains described by $m(x,y,\ell_d)=\sin(\pi x/\ell_d )$ and obtain $\ell_d=\alpha/\gamma$
for $H_r=H_e=0$ and neglecting the double well potential $V(m)$. Fig.\ref{figura2}(d) compares the domain width estimated from simulations at $h_e=0$ with the theoretical
estimate: the agreement implies that neglecting the anisotropy in the calculation appears to be a reasonable approximation. A better estimation has been obtained in Ref. \onlinecite{politi}.

A second important length-scale is given by the domain wall width $\ell_w$ which should induce a lower cutoff to the avalanche distribution.
If avalanche scaling is due to the dynamics of domain walls in a disordered medium then for lengthscales below $\ell_w$ the scaling should
break down since the very concept of domain wall is lost. Below this scale the inner structure of the domain wall and the rotation of the spins should become important for the avalanches, but these features can not be described by a simple scalar model. We estimate $\ell_w$  by energy minimization imposing a single domain wall at $x=0$. In the absence of external field and neglecting  dipolar interaction, Eq.(\ref{hamilton}) is minimized by a magnetization profile $m(x,y,\ell_w)=\tanh(x/\ell_w)$ with $\ell_w=\sqrt{2} \ell$ \cite{landau}. 
The domain wall width obtained in simulations is  
reported in Fig.\ref{figura2}(d) and compared with the theoretical estimate showing a good agreement at least for  large $\gamma$.

The third relevant lengthscale in the problem is the diameter of the nucleation droplet $\ell_n$. Bubble domains are stabilized by the dipolar interactions and $\ell_n$ has been estimated in ideal conditions \cite{aharoni,cape}. The theory, however, does not include disorder, 
leading to a overestimate of $\ell_n$. In fact, without disorder (i.e. for $D=0$) in the parameter range explored here, nucleation is completely
suppressed and magnetization reversal occurs coherently (i.e. uniformly in the entire sample, without domain nucleation)\footnote{In a real sample with very little disorder, magnetization reversal is known to occur by formation of stripes~\cite{saratz}, they typically nucleates at the sample edges, this cannot occur in our simulations due to periodic boundary conditions thus with $D=0$ the magnetization reverses all of the sudden.}. Hence, we conclude that the nucleation we observe is induced by disorder, but we were not able to estimate $\ell_n$ theoretically. 
In Fig.\ref{figura2}(d) we report the results of
numerical simulations showing that $\ell_n$ is slightly larger than $\ell_d$ and grows in a similar way with $\alpha/\gamma$.
Our theoretical estimates for the characteristic lengthscales compare nicely with the experimental results of Im et al. on CoCrPt films \cite{young} and of Schwarz et al. on LaSrMn0$_3$ films \cite{schwarz,liebmann}.  Using $K_u=2\times 10^{5}J/m^3$ and $A=3.8\times 10^{-12}$J/m for a $d=50$nm thickness CoCrPt film \cite{navas},  we estimate with our model $\ell_w=12 nm$, $\ell_d=28 nm$ and $\ell_n=36 nm$ while experiments yield $\ell_w=15 $nm, $\ell_d=30$nm and $\ell_n=40$nm. For a $d=100$nm thickness LaSrMn0$_3$ film \cite{liebmann}, we use $K_u=2\times 10^{4} J/m^3$ and $A=1.7\times 10^{-12} J/m$ and obtain $\ell_w=26$nm, $\ell_d=63$nm and $\ell_n=80$nm while in experiments we have $\ell_d=79$nm and $\ell_n=73$nm, again a very good agreement. 
\subsection{Avalanche statistics}
\begin{figure*}
\centering
\includegraphics[width=16.0cm,angle=0]{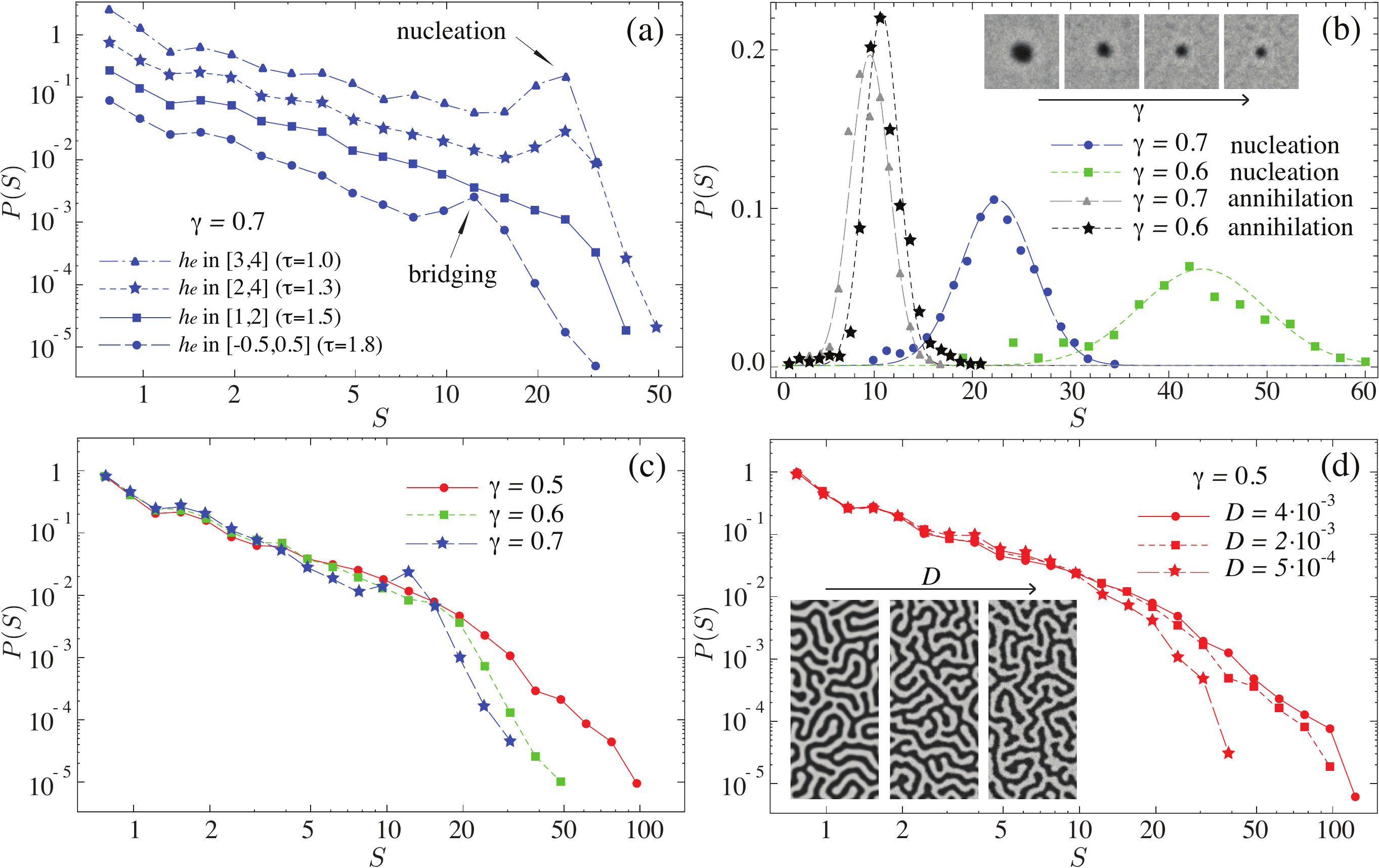}
\caption{Panel (a) shows the avalanche probability distribution for $\gamma=0.7$ measured in different regions of the hysteresis loop. The region is indicated in terms of the applied field $h_e$ interval together with the fitted power law exponents $\tau$. The curves are shifted to help comparison. Logarithmic binning has been adopted. Panel (b) shows the probability distribution of the areas of nucleated droplets and for the areas observed just before domain annihilation. The inset shows the size of the very first nucleated droplet for different simulations with growing $\gamma$. Panel (c) shows the avalanche size distribution for different $\gamma$ measured in the central region of the loop with $h_e$ ranging between $-0.5$ and $0.5$. Panel (d) shows the probability distribution of avalanche sizes for $\gamma=0.5$ at different disorder strength $D$. The inset highlights the change in the domain morphology induced by the disorder strengthening.}
\label{figura3}
\end{figure*}

The statistical analysis of the avalanche distribution is performed exactly as in  experiments \cite{schwarz,young}. We resolve single
avalanches by computing the difference between consecutive images of the magnetization and identifying the avalanche size $S$ as
the area of a connected cluster of reversed magnetization, defined by an appropriate threshold on the phase field (see also supplemental movies 5 and 6).
The probability distributions are obtained by logarithmic binning of the measured avalanche sizes, averaging over several 
realizations of the disorder field. All the measured distributions display a power law behavior $P(S)\propto S^{-\tau}$ for at least one decade
with the scaling regime limited at small and large sizes (see  Fig.\ref{figura3}).
Following Ref. \onlinecite{young}, we first compute the avalanche distribution for different values of the applied field along the hysteresis loop 
as shown in Fig.\ref{figura3}(a) for $\gamma=0.7$.  The four displayed distributions span a region ranging from the high fields where nucleation takes place to the center of the loop where bridging events appear, passing by intermediate regions where avalanches are mainly due to domain wall motion. We notice that the distribution upper cutoff decreases as we approach the center of the loop, where the domain density is maximum.
In most distributions, we observe a characteristic peak at large $S$ that is superimposed to the apparent power law decay. We suggest that 
these characteristic sizes are associated with nucleation events at small fields and bridging events at larger fields, since by removing those events we observe a marked decrease in the peaks amplitudes. A power law fit done excluding large values of $S$  
yields $\tau=1.0$ in the very first stage of the nucleation process when few isolated droplets expand freely, $\tau=1.3$ when the droplets increase their density, $\tau=1.5$ in the intermediate region and $\tau=1.8$ in the loop center. While we notice that the presence of a peak could strongly bias the estimate of the exponents, the values we get are in good agreement with the experimental results of 
Ref. \onlinecite{young} which report a similar trend in the variation of $\tau$ along the loop. 

Next, we distinguish between newly nucleated droplets and the expansion of existing domains and
measure the distributions of nucleated droplet sizes which follows a Gaussian distribution (see Fig.\ref{figura3}(b)) with 
a characteristic size decreasing with $\gamma$. The areas of annihilated domains are also Gaussian distributed but their typical sizes
is independent of $\alpha/\gamma$ and smaller than those of nucleated droplets. This is because, contrary to nucleation, 
the annhilation process does not involve an energy barrier. The domain size shrinks continuously until it becomes comparable to the domain wall width, when domains become unstable and disappear. Similar observations was made experimentally in Ref. \onlinecite{schwarz,liebmann}. 
In Fig.\ref{figura3} (c) we show the avalanche size distribution measured in the same central region of the hysteresis loop 
for different values of $\gamma$. The avalanche cutoff increases when $\gamma$ is decreased, following the corresponding increase of the domain width and confirming that $\alpha/\gamma$ is the relevant parameter controlling the size of the scaling regime, depicted by a shadowed region in Fig.\ref{figura2} (d). 

Finally Fig.\ref{figura3} (d) shows the effect of disorder strength on the avalanche size distribution in the central part of the
loop. The upper cutoff is found to increase with increasing disorder strength $D$. A similar result is found in domain wall depinning models \cite{zapperi98,durin00}, confirming that in this regime avalanches are due to domain wall motion. The inset of Fig.\ref{figura3}(d) refers to $\gamma=0.7$ and shows that, even if the domains shape is slightly affected by the disorder strength, $\ell_d$ is almost independent of $D$, while the nucleation diameter $\ell_n$ decreases with increasing $D$. This behavior has a simple explanation: to stronger disorder 
corresponds a larger value of the nucleation field $H_e$, but higher values of $H_e$ lead to a smaller nucleated droplet diameter $\ell_d$.
\section{Discussion}
In this paper, we have described the process of labyrinthine domain formation using a disordered phase-field model of magnetic
reversal in films with out-of-plane anisotropy. The model allows to overcome the limitations of spin models, where lattice
effects can effectively pin domain walls \cite{jagla1}, and at the same time provides an alternative to computationally more 
intense micromagnetic equations. We have modeled disorder using random-fields in analogy with other well studied hysteresis models \cite{sethna93},
but our main predictions should not change substantially with other forms of disorder. To check this, we have implemented also
random anisotropies \cite{jagla2} finding qualitatively similar results. The model successfully reproduces the main features observed experimentally
in out-of-plane magnetic thin films \cite{young,schwarz,liebmann}: nucleation of nearly circular domains, domain growth and branching
instabilities, labyrinthine domains expanding through avalanches and finally domain annihilation. Using simple arguments we estimate
the main characteristic lengthscales of the problem finding a good quantitative agreement with experiments. The domain wall and domain
widths, $\ell_w$ and $\ell_d$, are shown to delimit the scaling regime for the avalanches: the power law distribution of avalanches
is only observed between lower and upper cutoffs, determined by $\ell_w$ and $\ell_d$, respectively. Such a limited scaling regime
is expected to be the cause for the variability of the exponents reported in the literature for this class of materials. The competition
between dipolar and exchange interactions yield the observed labyrinthine domain structure and at the same time set strict 
boundaries for the scaling region. 

By measuring the avalanche size distribution along the hysteresis loop, we find effective exponents that vary considerably as
a function of the applied field in quantitative agreement with experimental results  for CoCrPt films \cite{young}. We tend to attribute
this variability in the exponents to a bias in the fit due to the upper and lower cutoff. The most reliable value we obtain is $\tau \simeq 1.5$ observed in the intermediate regime where nucleation and bridging events are not valid. This value is very close to the prediction of mean-field
theory that is found to be accurate for bulk samples \cite{zapperi98}, although we have no theoretical argument to justify it
for thin films. The values we find are, however, in net contrast with the one  
reported for LaSrMn0$_3$ film ($\tau \simeq 0.5$) \cite{schwarz,liebmann}. We notice, however, that the result is  due to
avalanches that occur for lengthscale smaller than the domain wall width $\ell_w$. For this reason, they can not be captured by
our model that does not involve spin rotation and other small scale details. Yet, we have reanalyzed the data of Ref. \onlinecite{schwarz,liebmann}
and found that on lengthscales larger than $\ell_w$ one sees a larger exponent although its determination is masked by the large
scale cutoff. 

The results of our numerical simulations yield some predictions that could be tested experimentally. Fig. \ref{figura3}(c) shows that 
the upper cutoff of the avalanche size distribution increases with $\gamma$. Experimentally it would be possible to tune $\gamma$
by analyzing magnetic films of different thickness. Similarly,  Fig. \ref{figura3}(d) reports a similar effect due to the strength of
the disorder, which could be changed experimentally by irradiating samples, as shown in Ref \cite{pilet}. Given the quantitative agreement
between our model estimates and the experimentally measured lengthscales, we are confident that this approach should be successful.\\

\section{Acknowledgements}
We thank G. Durin, A. Magni and J. P. Sethna for useful discussions. AB is supported by PRIN project 2008Y2P573.

\end{document}